\documentclass[12pt]{article}
\usepackage{graphicx}
\usepackage{cite}
\usepackage{amssymb,amsmath}
\begin{document}
\title{Positively and negatively hydrated counterions in molecular dynamics simulations of DNA double helix}%
\author{Sergiy Perepelytsya\\
Bogolyubov Institute for Theoretical Physics, NAS of
Ukraine,\\14-b Metrolohichna Str., Kiev 03143, Ukraine \\
perepelytsya@bitp.kiev.ua }\maketitle
\setcounter{page}{1}%
\maketitle
\begin{abstract}
The DNA double helix is a polyanionic macromolecule that in water solutions is neutralized by metal ions (counterions). The property of the counterions to stabilize the water network (positive hydration) or to make it friable (negative hydration) is important in terms of the physical mechanisms of stabilization of the DNA double helix. In the present research, the effects of positive hydration of Na$^{+}$ counterions and negative hydration of K$^{+}$ and Cs$^{+}$ counterions, incorporated into the hydration shell of the DNA double helix have been studied using molecular dynamics simulations. The results have shown that the dynamics of the hydration shell of counterions depends on region of the double helix: minor groove, major groove, and outside the macromolecule. The longest average residence time has been observed for water molecules contacting with the counterions, localized in the minor groove of the double helix (about 50 ps for Na$^{+}$, and lower than 10 ps for K$^{+}$ and Cs$^{+}$). The estimated potentials of mean force for the hydration shells of the counterions show that the water molecules are constrained too strong, and, consequently, the effect of negative hydration for K$^{+}$ and Cs$^{+}$ counterions has not been observed in the simulations. The analysis has shown that the effects of counterion hydration can be described more accurately with water models having lower dipole moments.
\end{abstract}
\section{Introduction}
The DNA macromolecule consists of two polynucleotide chains that in water solution are twisted around each other as a double helix \cite{Saenger}. The key feature of the structure as discovered in the famous research by James Watson and Francis Crick \cite{Watson} and supported by X-ray images by Rosalind Franklin \cite{Franklin} and Maurice Wilkins \cite{Wilkins}, is that the hydrophobic nucleotide bases are localized inside the double helix, minimizing the contacts with water molecules, while the backbone is faced to the solution. The backbone of the double helix comprises the phosphate groups, with each of them having a charge equal to -1\emph{e}. The negatively charged phosphate groups attract the positively charged ions from the solution resulting in the formation of a shell of counterions around the double helix \cite{Blagoy,Sivolob,Vologodskii}. The counterions neutralize the phosphate groups of DNA reducing the electrostatic repulsion of the opposite strands of the double helix. Therefore, water molecules and counterions of the ion-hydration shell around DNA are indispensable for the formation of the double helix and may be considered as integral part of its structure.

The physical properties of the DNA ion-hydration shell are essentially different from those of bulk water and depend on the region of the double helix: minor groove, major groove, and outside the macromolecule ~\cite{Maleev,Drew,Tereshko2,Mocci,Hynes2016,Hynes2017}. In the minor groove of the double helix, the mean residence time of water molecule is the longest (up to 100 ps), while in the major groove, especially, near the phosphate groups, it is several fold shorter than in the minor groove of the macromolecule \cite{Hynes2016,Hynes2017}. The counterions can disorder or stabilize the water structure inside the DNA ion-hydration shell. Depending on the structure organization of water molecules, the ions are usually classified into the positively hydrated and the negatively hydrated ones ~\cite{Ismailov}. In the case of the positively hydrated ions (Li$^{+}$, Na$^{+}$, and Mg$^{2+}$), the water molecules in the hydration shells are highly ordered, and the mean residence time of the molecule in the ion’s hydration shell is much longer than that in the bulk solution ~\cite{Ismailov}. Therefore, these ions are also known as the structure making ions \cite{Trostin}. In the case of the negatively hydrated ions (K$^{+}$, Rb$^{+}$, and Cs$^{+}$), the mean residence time of water molecules near the ion is shorter than that in the bulk ~\cite{Ismailov}, and the structure of the hydration shell is more friable than that of pure liquid water. Therefore, these ions are also known as the structure breaking ions \cite{Trostin}. The structure and dynamics of the DNA ion-hydration shell depend on the character of hydration of counterions.

To describe the structure of the ion-hydration shell of the DNA double helix, a theory similar to the statistical theory of electrolytes \cite{Yukhnovskii} shall be developed. In the present time, the features of the structure of counterion system around DNA are described within the framework of different theoretical approaches. In particular, the polyelectrolyte models that consider the macromolecule as a chain of charged beads or as a uniformly charged cylinder ~\cite{Manning, FK}, describe the effect of counterions condensation observed experimentally ~\cite{Das,Kwok,Andresen,Qiu}. On the other hand, considering the structure of DNA with counterions as an ionic type lattice (ion-phosphate lattice), the vibrations of counterions with respect to the phosphate groups have been found in the low-frequency Raman spectra of DNA ($<$ 200 cm$^{-1}$) ~\cite{PV1,PV2,BVKP, PV3,PV4,PV5}. The concept of the ion-phosphate lattice has been proven to be useful for describing different effects of DNA-counterion interaction ~\cite{Glibitski,Conductivity}. The existing theoretical approaches make a general outline of the structural and dynamical properties of the DNA-counterion systems.

In the theoretical descriptions the water around DNA is usually presented as a continuum with some dielectric constant. At the same time, for the consideration of the hydration effects of counterions the water shall be considered explicitly. In this regard, the method of classical molecular dynamics seems the most appropriate. The molecular dynamics studies ~\cite{Perepelytsya2018,Saba,Mocci,Pasi,Lavery2014,Liubysh,Canadian,SJAF} have shown that the counterion distribution around the double helix depend on the sequence of nucleotide bases, the region of the double helix, and are governed by the interplay between counterions and water molecules ~\cite{Pasi,Perepelytsya2018,Liubysh}. In particular, the study of counterion hydration \cite{Perepelytsya2018} has shown that the structure making ions interact mostly with DNA via water molecules of the hydration shell, while the structure breaking ions may squeeze through the DNA hydration shell to the groove bottom and form long-lived complexes with the atoms of nucleotide bases. The difference in the interaction of the structure making and the structure breaking counterions with the DNA double helix is explained by different structure and dynamics of the hydration shells of the ions.

The purpose of this research is to study the character of hydration of Na$^{+}$, K$^{+}$, and Cs$^{+}$ counterions incorporated into the hydration shell of the DNA double helix. To solve this problem, the trajectories of atomistic molecular dynamics simulations of DNA with counterions have been analysed. The radial distribution functions of water molecules with respect to the ions have been built, and the potentials of mean force have been derived. The average residence times of water molecules in the hydration shell of counterions have been estimated. The results have shown that the dynamics of the hydration shell of counterions depend on the region of the double helix, where the ion is localized. The positive character of hydration has been observed for all counterions. The effects of counterion hydration have been shown to be better described with the use of the water models having lower dipoles moments.

\section{Materials and methods}

The analysis of the structure and dynamics of the hydration shells of counterions, localized in different regions of the double helix, has been performed through molecular dynamics simulations \cite{Perepelytsya2018}. The simulations \cite{Perepelytsya2018} were carried out for the model systems of DNA in water solution with the counterions. The polynucleotide duplex d(CGCGAATTCGCG) that is known as  the Drew-Dickerson dodecamer \cite{Drew} was used as the model of DNA double helix. This fragment of DNA is characterized by the narrowed minor groove in the region with AATT nucleotide sequence (Fig. 1a). The major groove is visibly wider comparing to the minor groove. The DNA duplex was immersed into the water box 64$\times$64$\times$64 {\AA} with the metal ions of defined type: Na$^{+}$, K$^{+}$ or Cs$^{+}$. The number of counterions was 22 that was equal to the number of the DNA phosphate groups, making the system electrically neutral. As a result three systems of DNA water solution with the counterions of different type were studied: Na-DNA, K-DNA, and Cs-DNA.

The computer simulations \cite{Perepelytsya2018} were performed using NAMD software package \cite{Phillips} and CHARMM27 force field \cite{Foloppe,MacKerell}. The length of all bonds with hydrogen atoms was taken rigid using SHAKE algorithm \cite{SHAKE}. The TIP3P water model \cite{TIP3P} and the Beglov and Roux parameters of ions have been used \cite{Beglov}. The Langevin dynamics was used for all heavy atoms with the temperature 300 $^{\circ}$K. The total lengths of the simulation trajectory in the NVT ensemble  was more than 200 ns for each system. The simulation data were analyzed after 100 ns of equilibration of the systems. The details of the simulation process are described in \cite{Perepelytsya2018}.

\begin{figure}
\begin{center}
\resizebox{0.6\textwidth}{!}{%
  \includegraphics{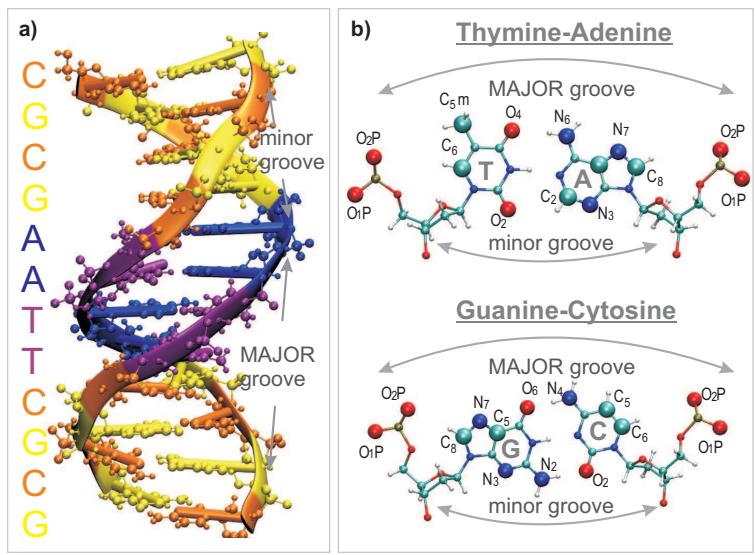}
}
\caption{a) The structure of the Drew-Dickerson dodecamer d(CGCGAATTCGCG) \cite{Drew}. The minor and major grooves are indicated. Nucleotide colour scheme: Cytosine (orange), Guanine (yellow), Adenine (blue), Thymine (purple). b) Thymine-Adenine and Guanine-Cytosine nucleotide pairs and the reference atoms shown as enlarged spheres with the names of atoms that were used as reference points for the construction of the radial distribution functions.}
\label{fig:twist}       
\end{center}
\end{figure}

In the present work the VMD software \cite{Humphrey} was used for the analysis and visualization. Using the VMD plug-in \cite{gr}, the radial distribution functions (RDFs) have been calculated by the  following formula:
 \begin{equation}\label{Eq0}
g(r)=p(r)(4\pi{r^2}\Delta{r}N_{p}/V),
\end{equation}
were $p(r)$ is the average number of atomic pairs, found at the distance within $(r\div{r+\Delta{r}})$; $N_{p}$ is the number of pairs of selected atoms; $V$ is the total volume of the system; $\Delta{r}$  is the width of histogram bins which in the present work was taken equal to 0.5 \AA . The average number of atomic pairs has been calculated every 10000 time steps that is 500 frames per the nanosecond.

The RDFs have been built for oxygen atoms of water molecules with respect to the ions localized in different regions of the double helix: in the minor and major grooves (RDF$_{Ion}^{minor}$ and RDF$_{Ion}^{major}$), near the phosphate groups (RDF$_{Ion}^{ph}$), and in the bulk (RDF$_{Ion}^{bulk}$). The counterion has been considered to be localized in some region of the double helix if it was within 5 {\AA} of one of the reference atoms. The reference atoms of DNA are shown on the Figure 1b and indicated in the Table 1. The same radial distribution functions have been calculated for water molecules with respect to other water molecules, localized in different regions of DNA macromolecule (RDF$_{W}^{minor}$, RDF$_{W}^{major}$, RDF$_{W}^{ph}$, and RDF$_{W}^{bulk}$). The reference water molecules were not in direct contact with the atoms of the DNA macromolecule.

\begin{table}[!]
\noindent\caption{The reference atoms of DNA macromolecule for the radial distribution functions.}\vskip3mm\tabcolsep4.5pt
\noindent{\footnotesize
\begin{tabular}{lcccc}
 \hline%
 \multicolumn{1}{c}{\rule{0pt}{5mm} DNA region}%
 & \multicolumn{1}{c}{ Adenine}
 & \multicolumn{1}{c}{ Guanine}
 & \multicolumn{1}{c}{ Thymine}
 & \multicolumn{1}{c}{ Cytosine}\\[2mm]%
\hline%
\rule{0pt}{5mm}Minor groove&N$_{3}$, C$_{2}$&N$_{3}$, N$_{2}$&O$_{2}$&O$_{2}$\\ 
Major groove&N$_{6}$, C$_{5}$,& O$_{6}$, C$_{5}$,& O$_{4}$, C$_{5}$m, &N$_{4}$, C$_{5}$, \\%
&N$_{7}$, C$_{8}$& N$_{7}$, C$_{8}$& C$_{6}$&C$_{6}$\\%
Phosphates&PO$_{1,2}$&PO$_{1,2}$&PO$_{1,2}$&PO$_{1,2}$\\[2mm]%
\hline
\end{tabular}
}
\end{table}

\section{Results}

\emph{Radial distribution functions.} The obtained  radial distribution functions of water molecules with respect to the counterions (ion-water RDFs) are characterized by two maximums: the first is intensive and the second is weak (Fig. 2a). The position of maximums are governed by the size of counterion and water molecule. The intensities of the first and the second maximums depend on a region of the double helix where the counterion is localized. The only exception was observed in the case of the first maximum for ion-water RDFs of Na$^{+}$ counterions that have approximately the same hight for all  regions of counterion localization. In the same time, in the case of K$^{+}$ and Cs$^{+}$ counterions the difference is essential in the case of the both the first and the second maximums.

\begin{figure}
\begin{center}
\resizebox{0.6\textwidth}{!}{%
  \includegraphics{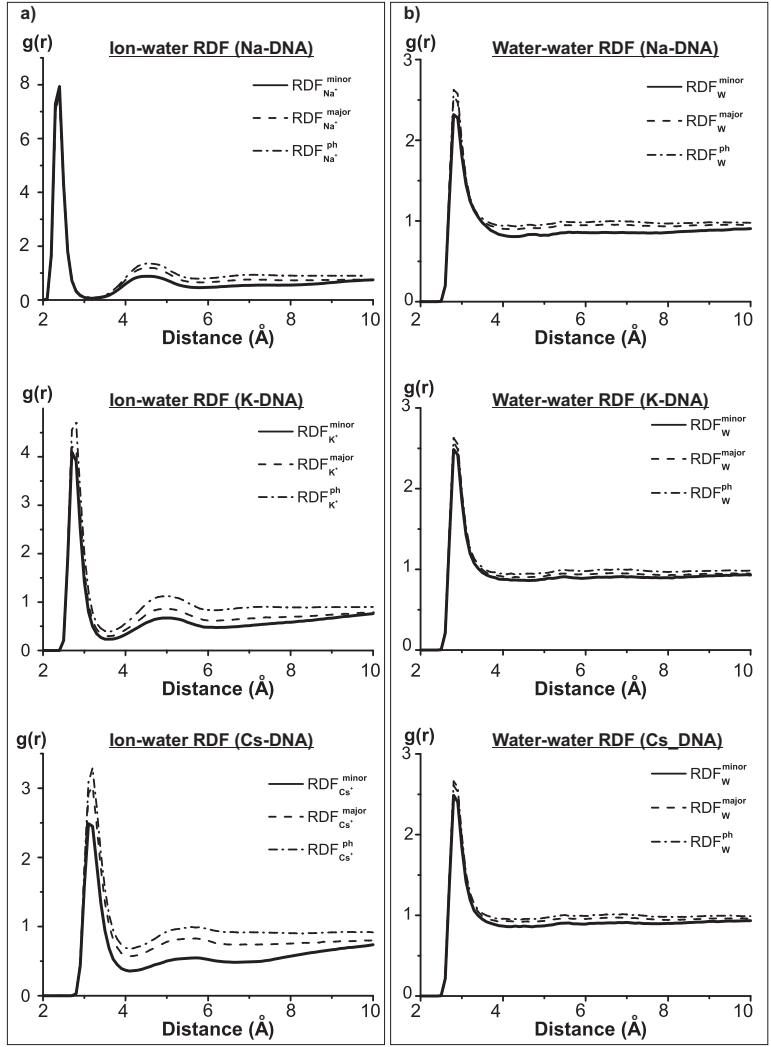}
}
\caption{The radial distribution functions (RDFs) for the oxygen atoms of water molecules with respect to Na$^{+}$, K$^{+}$, Cs$^{+}$ counteions (a) and with respect to the other oxygen atoms of water molecules (b) in different regions of the double helix: minor groove (RDF$_{Ion}^{minor}$ and RDF$_{W}^{minor}$), major groove  (RDF$_{Ion}^{major}$ and RDF$_{W}^{major}$), and near the phosphate groups of DNA backbone (RDF$_{Ion}^{ph}$ and RDF$_{W}^{ph}$).}
\end{center}
\end{figure}

The RDFs of water molecules with respect to water molecules (water-water RDFs) are characterized by the strong first maximum and flat curve after (Fig. 2b). The second maximum is very weak and hardly visible. The obtained shapes of the RDFs are characteristic for the TIP3P water model \cite{TIP3P,modelRDF}. The difference between water-water RDFs for the case of different regions of the double helix is observed only for the first peak that has always lower intensity in the case of water molecules in the minor groove.

\emph{Potential of mean force.} A water molecule in the hydration shell of an ion is trapped in the potential well and separated from the outer water layer by the potential barrier (Fig. 3). In the present work the potential barriers for water molecules were estimated using the potentials of mean force (PMF) derived from the radial distribution functions:
\begin{equation}\label{Eq1}
E(r)=-k_{B}T\ln{(g(r))},
\end{equation}
$k_{B}$ is the Boltzmann constant, $T$ is the temperature.

\begin{figure}
\begin{center}
\resizebox{0.6\textwidth}{!}{%
  \includegraphics{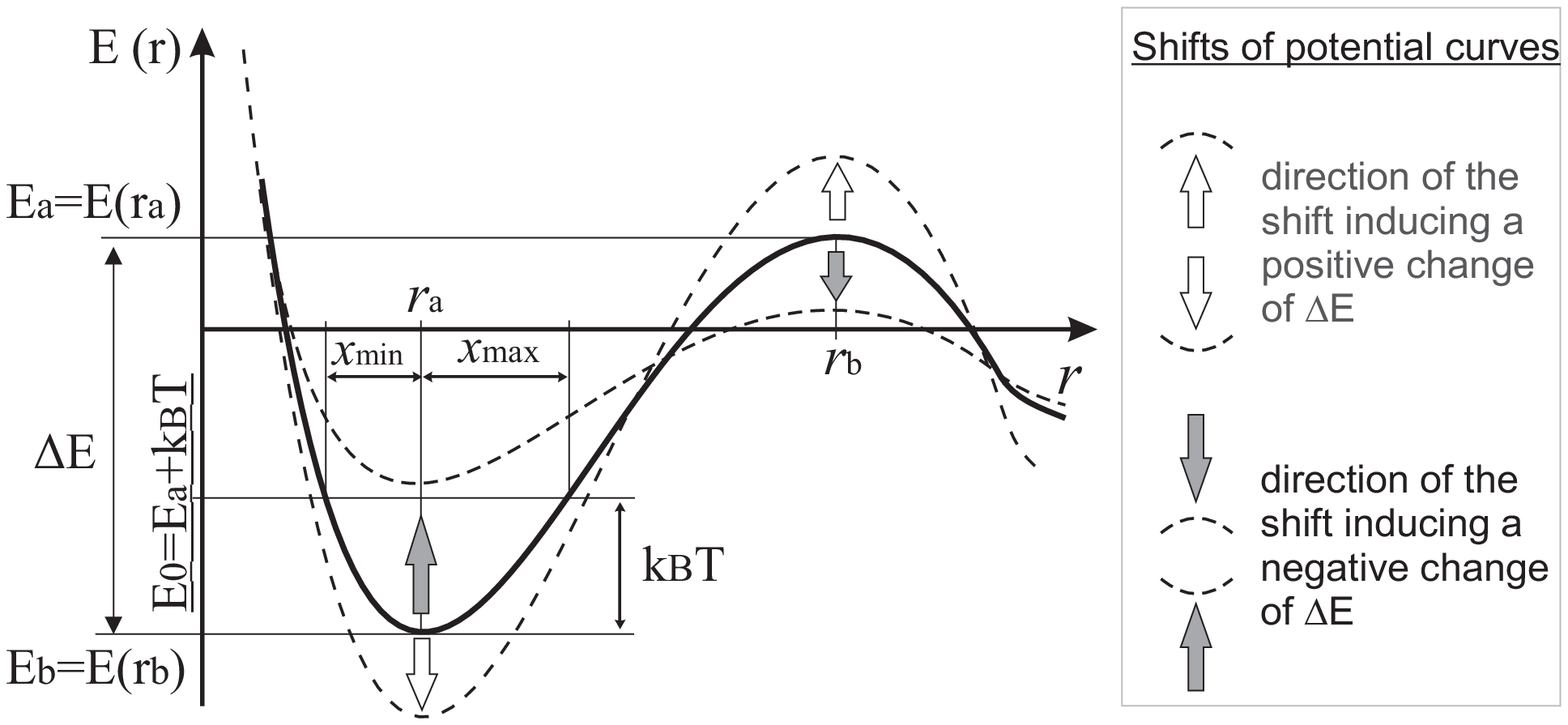}
}
\caption{The schematic structure of mean field potential energy of water molecule, derived from a radial distribution function. $\Delta{E}$ is the potential energy barrier; $E_a$ and $E_b$ are the energy values in the minimum ($r_a$) and maximum ($r_b$), respectively.}
\end{center}
\end{figure}

The calculated potentials of mean force are characterized by two potential wells (Fig. 4). In the present work the dynamics of water molecule in the first hydration shell is in the scope of interest, therefore the first potential well and the potential barrier ($\Delta{E}$) between the first and the second potential wells have been studied.

\begin{figure}
\begin{center}
\resizebox{0.6\textwidth}{!}{%
  \includegraphics{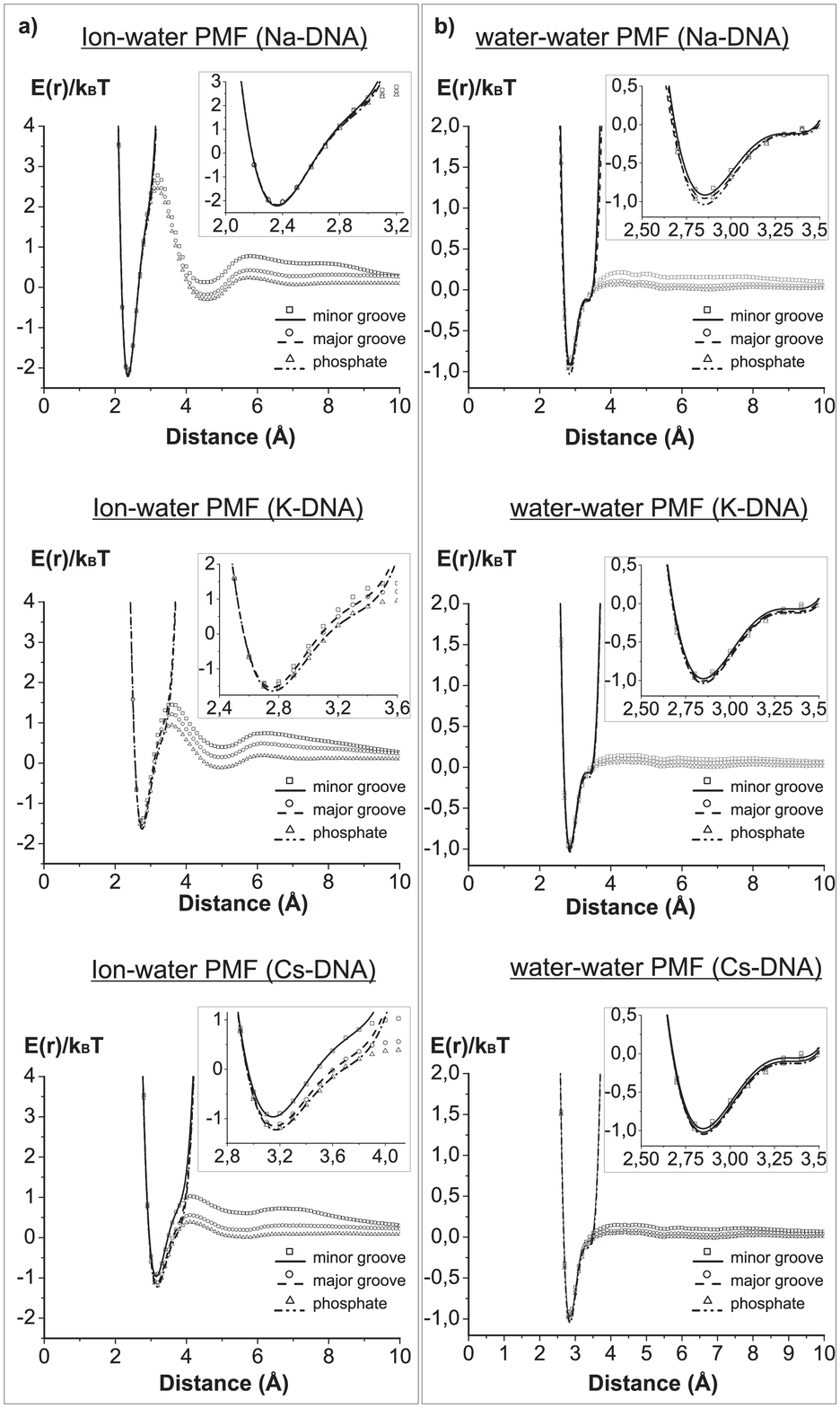}
}
\caption{The potentials of mean force of water molecules (PMFs). a) The PMFs for water molecules with respect to Na$^{+}$, K$^{+}$, Cs$^{+}$ in different regions of the double helix. b) The PMF for water molecules with respect to the oxygen atoms of water molecules in different regions of the double helix. The The lines correspond to curves fitted to PMF: solid, dashed and dotted lines correspond to the cases of the counterion minor groove, in the major groove, and near the phosphate group of DNA, respectively.}
\end{center}
\end{figure}

The results show that ion-water PMFs have different shape and depth in the case of different counterions. In the case of Na$^{+}$ the potential well is the deepest, while in the case of Cs$^{+}$ it is the smallest (Fig. 4a). The difference of ion-water PMFs is also observed for different regions of the double helix, where the counterion may be localized. In the same time, the water-water PMF are rather similar, and the difference is hardly visible for different regions of the double helix.

Using the obtained PMFs, the parameters describing the energy of counterion hydration were calculated by the formula (\ref{Eq1}). The resulted values of the potential barriers for water molecule in the hydration shell of the ion ($\Delta{E}_{ion}$) are the highest in the case of Na$^{+}$ ions, while in the case of K$^{+}$ and Cs$^{+}$ ions the values of $\Delta{E}_{ion}$ are about two times lower. Such behaviour is the result of different surface charge of the ions due to their different size. The energy barrier for water molecule in the hydration shell of the counterion is much higher than the energy barrier of water molecule in the bulk ($\Delta{E}_{ion}>\Delta{E}_{w}$), moreover in the case of counterions in the minor groove the potential barrier  is the highest, while in the case of counterion near the phosphate groups and in the bulk it is much lower (Table 2).

\begin{table}
\noindent\caption{ Parameters of the potentials of mean force. The positions of minimum and maximum of the potential well $r_{a}$ and $r_{b}$, respectively (in \AA). The values of the potential barriers in kcal/mol for water molecules in the hydration shell of the ion ($\Delta{E}_{ion}$) and for water molecule surrounded by the other water molecules ($\Delta{E}_{w}$).}\vskip3mm\tabcolsep4.5pt
\noindent{\footnotesize
\begin{tabular}{lccc}
 \hline%
 \multicolumn{1}{l}{\rule{0pt}{5mm} System}%
 & \multicolumn{1}{c}{ Na-DNA }
 & \multicolumn{1}{c}{ K-DNA  }
 & \multicolumn{1}{c}{ Cs-DNA }\\[2mm]%
                                           \hline%
      \textbf{Ion-water}&$r_{a}$ $r_b$ $\Delta{E}_{ion}$&$r_{a}$ $r_b$ $\Delta{E}_{ion}$&$r_{a}$ $r_b$ $\Delta{E}_{ion}$\\[2mm]%
               Minor gr.&2.36 3.20 3.00&2.75 3.60 1.77&3.15 4.10 1.20\\ %
               Major gr.&2.36 3.20 2.85&2.75 3.60 1.65&3.16 4.10 1.03\\%
                 Phosph.&2.36 3.20 2.78&2.75 3.60 1.45&3.17 4.10 0.97\\%
                    Bulk&2.32 3.15 2.82&2.71 3.55 1.58&3.10 4.05 1.03\\[2mm]%
\hline
    \textbf{Water-water}&$r_{a}$ $r_b$ $\Delta{E}_{w}$&$r_{a}$ $r_b$ $\Delta{E}_{w}$&$r_{a}$ $r_b$ $\Delta{E}_{w}$\\[2mm]%
               Minor gr.&2.85 4.20 0.68&2.85 4.20 0.67&2.85 4.20 0.68\\ %
               Major gr.&2.85 4.30 0.64&2.85 4.20 0.67&2.85 4.40 0.66\\%
                 Phosph.&2.85 4.25 0.67&2.85 4.30 0.66&2.85 4.20 0.66\\%
                    Bulk&2.79 4.05 0.67&2.84 4.20 0.67&2.84 4.20 0.69\\[2mm]%
                    \hline
\end{tabular}
}
\end{table}

The calculated values of the potential barriers (Table 2) have been compared with the results of molecular dynamics simulations \cite{Koneshan} for the ions in aqueous solutions at 25 $^{\circ}$C. The values of $\Delta{E_{ion}}$ for water molecules in the hydration shell of  Na$^{+}$, K$^{+}$, and Cs$^{+}$ ions obtained in the work \cite{Koneshan} are equal to 2.3 kcal/mol, 1.3 kcal/mol, and 0.9 kcal/mol, respectively. Such values are rather close to our results, but in general the potential barriers in the Table 2 are slightly higher than in the work \cite{Koneshan}. The reason of different values of $\Delta{E_{ion}}$ may be related to different water models and ion parameters that have been used in the present work and in the work \cite{Koneshan}.

The difference of the energy barriers for water molecule in the hydration shell of counterion and in the bulk ($d{E}=\Delta{E_{ion}-\Delta{E_w}}$) determines the character of counterion hydration. The structure making (positively hydrated) ions have $d{E}>0$, while the structure breaking  (negatively hydrated) ions are characterized by $d{E}<0$. The obtained results (Table 2) show that the values of $d{E}$ are positive for all counterions: 2.15 kcal/mol, 0.91 kcal/mol, and 0.34 kcal/mol for Na$^{+}$, K$^{+}$, and Cs$^{+}$ counterions, respectively. In the same time, the experimental data \cite{Ismailov} show that among the considered counterions only sodium ion has the positive difference of the potential barriers $d{E}=0.25$ kcal$/$mol, and its is much lower than in the simulations. The potassium and cesium ions have a negative difference of the potential barriers $d{E}=-0.25$ kcal$/$mol and $d{E}=-0.33$ kcal$/$mol respectively, which is why they are negatively hydrated \cite{Ismailov}. The reason of the difference between obtained energy barriers and experimental data may be related to the parametrization of water models that will be discussed in the following section.

\emph{Residence time.} The determined potentials of mean force allows us to estimate the residence time $\tau$ for water molecules using the equation of Arheniuns type \cite{Ismailov}. In the present work this equation is presented in the following form:
\begin{equation}\label{Eq3}
\tau=2\tau_{0}\exp{\left(\frac{\Delta{E}}{k_{B}T}\right)},
\end{equation}
where $\tau_{0}$ is the characteristic time of approaching of the molecule to the potential barrier $\Delta{E}$. The factor 2 in the formula (\ref{Eq3}) means that the water molecule being at the top of the potential barrier may leave the hydration shell or return back to the potential well with the equal probability. The value of $\tau_{0}$ is estimated as a period of vibrations of a particle in a potential well using the law of energy conservation for the finite motion \cite{Landau}:
\begin{equation}\label{Eq4}
  \tau_{0}=\sqrt{2\mu}\int_{x_{min}}^{x_{max}}{\frac{dx}{\sqrt{E_0 - E(x)}}},
\end{equation}
where $\mu$ is the mass of a water molecule; $x=r-r_a$ is the displacement of a water molecule from equilibrium position $r_a$; $x_{min}$ and $x_{max}$ are the amplitude displacements; $E_0$ is the amplitude energy of vibrations of a water molecule in the potential well (Fig. 3). The potential function $E(x)$ is determined from the potential of mean force in approximation of polynomial function:
\begin{equation}\label{Eq5}
E(x)\approx{ E_a + C_2x^2/2 + C_3x^3/3 + C_4x^4/4 },
\end{equation}
where $E_a$ is the depth of the potential well; $C_2$, $C_3$, $C_4$ are the fitting parameters. The amplitude displacement ($x_{min}$ and $x_{max}$) were determined from the condition: $E(x)=E_0$ (Fig. 3). Taking into account the Boltzmann's law of equal distribution of energy over degrees of freedom, the magnitude of the vibrational amplitude energy was determined as follows: $E_0=E_a + k_{B}T$. By substituting (\ref{Eq5}) to the equation (\ref{Eq4}) the elliptic integral was obtained.

\begin{table}
\noindent\caption{ The residence times ($\tau$) and the half-period of vibration ($\tau_{0}$) in ps for water molecules in the hydration shell of counterion and  surrounded by other water molecules. }\vskip3mm\tabcolsep4.5pt
\noindent{\footnotesize
\begin{tabular}{lccc}
 \hline%
 \multicolumn{1}{c}{\rule{0pt}{5mm} System}%
 & \multicolumn{1}{c}{ Na-DNA }
 & \multicolumn{1}{c}{ K-DNA  }
 & \multicolumn{1}{c}{ Cs-DNA }\\[2mm]%
 \hline%
 Ion-water&$\tau$  $\tau_{0}$&$\tau$  $\tau_{0}$&$\tau$  $\tau_{0}$\\ %
               Minor gr.&52.33  0.18&8.87  0.23&5.02 0.34\\ %
               Major gr.&41.54  0.18&7.60  0.24&4.29  0.39\\%
                 Phosph.&37.02  0.18&5.26  0.23&4.10 0.41\\%
                       Bulk&39.28  0.18&6.51  0.24&4.05  0.36\\[2mm]%
 \hline
 Water-water&$\tau$  $\tau_{0}$&$\tau$  $\tau_{0}$&$\tau$  $\tau_{0}$\\ %
               Minor gr.&3.61  0.58&4.22  0.69&4.34 0.71\\ %
               Major gr.&3.79  0.65&4.19  0.69&4.31  0.72\\%
                 Phosph.&4.15  0.69&4.23  0.71&4.68 0.85\\%
                       Bulk&3.25  0.54&3.29  0.54&4.86  0.74\\[2mm]%
\hline
\end{tabular}
}
\end{table}

The calculated average residence times of water molecules in the hydration shell of the ion are within the range from about 2 ps to 50 ps (Table 3). The longest residence time was observed for the case of sodium counterions, while in the case of potassium and cesium ions it is several times lower. The dependence of $\tau$ values on the region of counterion localization is also observed. The longest residence times of water molecules have been observed for the hydration shells of the ions localized in the minor groove  of the double helix ($\tau_{minor}$), while in the major groove ($\tau_{major}$), and near the phosphate group of the macromolecule backbone ($\tau_{ph}$) the residence times are shorter: $\tau_{minor}>\tau_{major}>\tau_{ph}$.

The residence time of water molecules near the counterions has been also calculated using the simulation trajectories. In the analysis, a water molecule was taken into account as part of hydration shell of the ion if the oxygen atom was localized within the distance $r_h$. The distance $r_h$ characterizes the outer boundary of the hydration shell and should be close to the minimum of the radial distribution function (the values $r_b$ in the Table 2). The values of the residence times were averaged over all counterions of the system. The obtained values of the residence times increase as the  distance $r_h$ increases (Fig. 5). In the case of Na$^+$ counterions the $\tau$ values obtained from PMF (Table 3) agree with the values obtained from the analysis of simulation trajectories at the distance $r_h=r_b$ (grey region on the Figure 5). In the case of K$^+$ and Cs$^+$ counterions the values of the residence times calculated by different methods are close at some distance that is lower than  $r_b$. The difference $r_h-r_b$ is less than 0.2 {\AA} that is within the accuracy limit of the estimations. The $\tau$ values for water molecules in the bulk (inset on Figure 5) are in agreement with the calculated values from PMF. Thus, the values of the average residence times of water molecules, estimated in the present work by two different methods, are in the sufficient agreement.

\begin{figure}
\begin{center}
\resizebox{0.6\textwidth}{!}{%
  \includegraphics{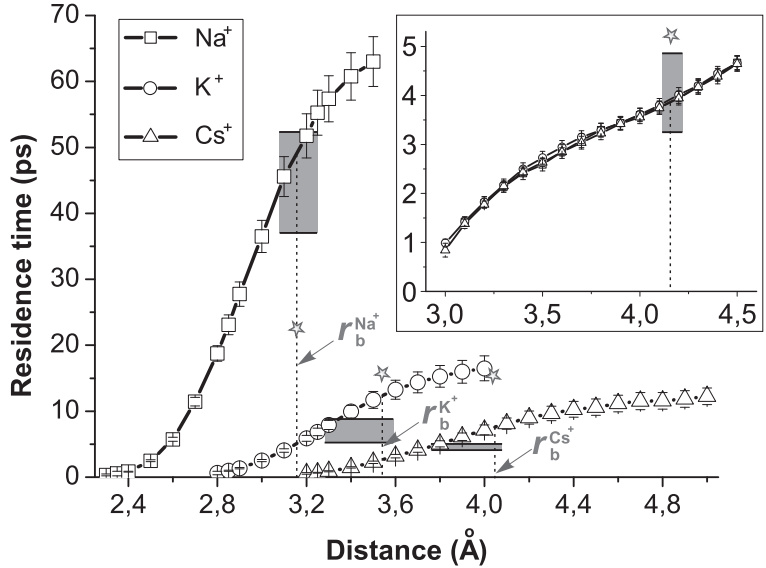}
}
\caption{The dependence of the average residence time of water molecules near Na$^{+}$, K$^{+}$, Cs$^{+}$ counterions on the distance from the ion. The range of residence times from Table 3 is grayed out. The dotted lines shows the distances of RDF minima ($r_b$) that also correspond to the position of the potential barriers of PMF. Inset: the dependence of the average residence time on  O-O distance for water molecules in the bulk in the case of three simulated systems. The asterisks show the values of the residence times obtained in the work \cite{Koneshan}.}
\end{center}
\end{figure}

The obtained average residence times (Table 3) have been compared with the other results of molecular dynamics simulations for alkali metal ions in water solutions \cite{Koneshan}. The residence times obtained in the work \cite{Koneshan} have the same dependence on the ion size  as in the present work, but the absolute $\tau$ values are different (asterisks on the Figure 5). There may be several reasons of such disagreement.  However, the most essential reason may be the use of  different definitions of the residence times and the methods for its calculation. In the present work the residence times was calculated from the mechanistical approximation of the motion of water molecule in the potential well that was obtained on the basis of the potential of mean force. In the work \cite{Koneshan} the residence time was calculated as the integral of the time correlation functions. These methods are not equivalent and the additional analysis should be done to find where they give the same result.

Thus, the results of molecular dynamics simulations for DNA with the positively hydrated Na$^{+}$, and negatively hydrated K$^{+}$ and Cs$^{+}$ counterions  show that the dynamics of water molecules in the hydration shells of counterions depends on their localization around the double helix. In particular, the longest residence time was observed for a water molecule near the counterion that is localized inside the minor groove of the double helix, and it is longer than for the case of a water molecule near the same ion but in a bulk water. This difference may be due to the confined space inside the double helix and due to the structured system of water molecules that is formed in DNA grooves. In the same time, the results clearly show that the  obtained energy barriers for water molecules near the ions are too high, making the hydration shell too rigid. The counterions Na$^{+}$, K$^{+}$, and Cs$^{+}$ in the simulated systems are positively hydrated, and the effect of negative hydration for K$^{+}$, and Cs$^{+}$ was not observed.

\section{Discussion}
To explain the reason of high values of the potential barriers the possible influence of water model should be analyzed. The TIP3P water model that was used in the simulations is characterized by the dipole moment value 2.35 D, while the experimental value for water molecule in gas phase is 1.86 D and in liquid phase is 2.95 D \cite{Gubskaya}. In this regard, let us analyze the potential barrier as a function of dipole moment. For this purpose the potential of mean force has been estimated as the change of free energy of water molecule after its replacement from the hydration shell of the ion to the bulk water:
\begin{equation}\label{Eq7}
  E\equiv\Delta{G}=\Delta{H}-T\Delta{S}+\Delta{G_0},
\end{equation}
where $\Delta{H}$ and $T\Delta{S}$ are the enthalpy and entropy contributions, and $\Delta{G_0}$ is some constant part of the free energy change.

The enthalpy contribution is featured mostly by the interaction of water molecule with the ion. In the work ~\cite{Biakov} the energy of water molecule near the ion was successfully described by presenting the water molecule as a dipole in the field of the ion. In our model the repulsion between water molecule and ion at small distances is also taken into consideration. As the result the enthalpy change may be presented as a sum of average dipole-dipole ($\overline{U_{i-d}(r)}$) and repulsion ($\overline{U_{rep}(r)}$) terms:
\begin{equation}\label{Eq8}
  \Delta{H}=\overline{U_{i-d}(r)}+\overline{U_{rep}(r)}.
\end{equation}

Taking into consideration that the direction of a dipole vector in the electric field of the ion is described by the Boltzmann distribution an average ion-dipole interaction may be presented in the following form:
\begin{equation}\label{Eq12}
\overline{U_{i-d}\left(r\right)}=-k_BTL(\alpha){\alpha(r)},
\end{equation}
where $L(\alpha)=\coth{\alpha}-\alpha^{-1}$ is the Langevin function, and
\begin{equation}\label{Eq10}
\alpha(r)=\frac{1}{k_BT}\cdot\frac{qd}{4\pi\varepsilon\varepsilon_{0}r^2}.
\end{equation}
Here $q$ is the charge of the ion; $\varepsilon$ is the dielectric constant of the media near the ion; $\varepsilon_{0}$ is the dielectric constant of vacuum; $d$ is the dipole moment of water molecule.

The repulsion between water molecule and ion is described by the potential in Born-Mayer form that is used in the description of ions in ionic crystals \cite{Kittel} and DNA ion-phosphate lattice \cite{PV4}:
\begin{equation}\label{Eq13}
  U_{rep}\left(r\right)= Ae^{-r/b},
\end{equation}
where $A$ and $b$ are the parameters describing repulsion between ion and water molecule as hard cores.

The motions of dipole moments of water molecules around the ion are hindered and in general the molecules are highly oriented due to the electrostatic field. Therefore, in our model the entropy is assumed to increase with ion-water distance the same as the average direction of water dipole, described by the Langevin function $L(\alpha)$. As a result the change of entropy is presented as follows:
\begin{equation}\label{Eq14}
    \Delta{S}=-s_0L(\alpha),
\end{equation}
where $s_0$ is the entropy of water molecule in the bulk. The parameters $A$ and $s_0$  we derive from the condition for maximum and minimum at the distances  $r_a$ and $r_b$: $\frac{d\Delta{G}}{dr}|_{r=r_a}=0$, $\frac{d\Delta{G}}{dr}|_{r=r_b}=0$.

The energy contribution $\Delta{G_0}$ is featured by the interaction energy with other water molecules of the system that includes the both enthalpy and entropy contributions. The determination of this contribution is a complex problem and it is not necessary for the estimation of the potential barrier hight. In the present work it is determined from the condition $\Delta{G(r_c)}=0$, here $r_c$ is some point,  where the potential of mean is equal to zero. Taking this into consideration and using the equations (\ref{Eq7}) -- (\ref{Eq16}), the change of the potential of mean force in $k_BT$ units may be written in the following from:
\begin{equation}\label{Eq16}
\widetilde{E}(r)=-L(\alpha)[\alpha(r)+s_0]+Be^{-\frac{r-r_a}{b}}+\Delta{g_0},
\end{equation}
where $\widetilde{E}(r)=E/k_BT$, $B=-A/k_BT$; $\Delta{g_0}=-L(\alpha_c)(\alpha_c+s_0)+Be^{-(r_c-r_a)/b}$, and $\alpha_c=\alpha(r_c)$.

To estimate the energy using the formula \ref{Eq14} the parameters have been determined in the following way. The repulsion parameter is taken the same as in the case of the crystals of alkali metal ion that is $b\approx0.3$\AA. The temperature is taken the same as in the molecular dynamics simulations $T=300^{\circ}$K. The dipole moment  $d=2.35$D was taken the same as in TIP3P model of water molecule. The values of equilibrium distances ($r_a$) and the barrier distance ($r_b$) were taken from the Table 2. The distance $r_c$ is defined as: $r_c=(r_a+r_b)/2$. The dielectric constant has been determined using the dielectric function \cite{Lavery}, developed for the description of the electrostatic interactions in nucleic acids: $\varepsilon(\widetilde{r})=78-77(0.0128\widetilde{r}^2+0.16\widetilde{r}+1)e^{-0.16\widetilde{r}}$,
where $\widetilde{r}$ is the distance between charges in Angstroms. At the distance about $(2\div4)$  {\AA} this function gives the value within the range $\varepsilon\approx(1.3\div3)$.

\begin{figure}
\begin{center}
\resizebox{0.6\textwidth}{!}{%
  \includegraphics{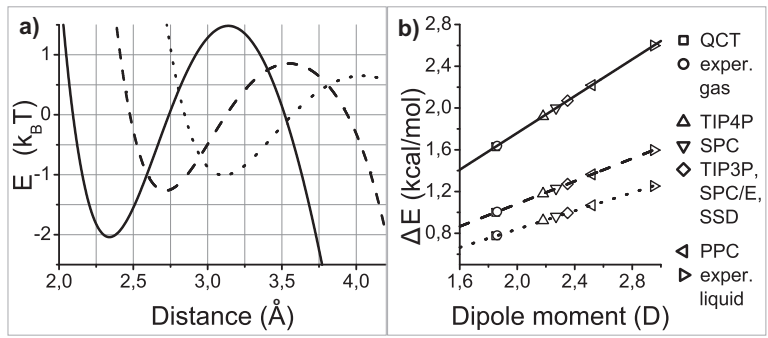}
}
\caption{a) The distance dependence of the energy of water molecule in the first hydration shell of the ion. b) The dependence of the potential barrier $\Delta{E}$ on the dipole moment of water molecule. The values of the potential barriers that correspond to the dipole moments of different water models \cite{QCT,TIP4P,SPCE,SSD,SPC}  and experimental values for water molecule \cite{Gubskaya} are shown as the figured points. The lines on the both (a) and (b) figures made in solid, dashed, and dotted style correspond to Na$^{+}$, K$^{+}$, and Cs$^{+}$ ions, respectively.}
\end{center}
\end{figure}

The estimations show that the hight of the potential barrier $\Delta{E}=\widetilde{E}(r_b)-\widetilde{E}(r_a)$ decreases as the size of the counterion increases: $\Delta{E_{Na^+}}>\Delta{E_{K^+}}>\Delta{E_{Cs^+}}$ (Fig. 6a). The same dependence of $\Delta{E}$ has been observed in our molecular dynamics simulations (Figure 4 and Table 2). To analyse the role of dipole moment of water molecule the potential barrier has been calculated using the formula (\ref{Eq16}) for different values of dipole moment (Fig.  6b). The results show that the potential barrier for water molecule in the hydration shell of the ion increases linearly as the value of dipole moment increases. The values of the potential barriers that correspond to the  dipole moments of different  water models and experimental data \cite{Gubskaya,QCT,TIP4P,SPCE,SSD,SPC} (the points on the Figure 6b)  are within the range $(1.6\div2.6)$ kcal/mol, $(1.0\div1.5)$ kcal/mol, and $(0.7\div1.2)$ kcal/mol for Na$^{+}$, K$^{+}$, and Cs$^{+}$ ions, respectively. Taking this into consideration it may be concluded that the models of water molecule with lower dipole moments should give more accurate description of the hydration effects of counterions.

\section{Conclusions}
The dynamics of water molecules in the hydration shell of the positively (Na$^{+}$) and negatively (K$^{+}$ and Cs$^{+}$) hydrated counterions around the DNA double helix has been studied. The molecular dynamics simulations for the DNA fragment in water solution with the counterions under the temperature 300$^{\circ}$ K have been used. As the result the potential barriers and the residence times of water molecules near the counterions have been estimated. The analysis show that the dynamics of water molecules in the hydration shell of counterions depends on their localization around the double helix that is the manifestation of the interplay between water molecules in the hydration shell of DNA and counterions. The longest residence time of water molecule has been observed for the case of counterions in the minor groove of the double helix: about 50 ps for Na$^{+}$ counterion and lower than 10 ps for K$^{+}$ and Cs$^{+}$ counterions. In the major groove and outside the double helix it is essentially lower. In the simulations the counterions constrain water molecules too strong making the hydration shell more rigid than it should be. As the result the effect of negative hydration in the case of K$^{+}$ and Cs$^{+}$ counterion was not obtained. The analysis, performed within the framework of the developed phenomenological model, has been showed that the strength of the hydration shell is proportional to the value of dipole moment of water model. The water models with lower dipole moments are expected to give better description of the effects of counterion hydration.

\vskip3mm \textit{Acknowledgement.}
The present work was partially supported by the Project  of the Department of Physics and Astronomy of the National Academy of Sciences of Ukraine (0117U000240).

\end{document}